\documentclass[%
 reprint,
 amsmath,amssymb,
 aps,
]{revtex4-1}

\usepackage{graphicx}
\usepackage{dcolumn}
\usepackage{bm}
\usepackage{hyperref}

\begin{document}

\preprint{APS/123-QED}

\title{DeepEfficiency - optimal efficiency inversion in higher dimensions at the LHC}

\author{Mikael Mieskolainen}
\email{mikael.mieskolainen@cern.ch}
\affiliation{
Department of Physics, University of Helsinki and Helsinki Institute of Physics, P.O. Box 64, FI-00014 Helsinki, Finland}


\date{September 16, 2018}

\begin{abstract}
We introduce a new high dimensional algorithm for efficiency corrected, maximally Monte Carlo event generator independent fiducial measurements at the LHC and beyond. The approach is driven probabilistically using a Deep Neural Network on an event-by-event basis, trained using detector simulation and even only pure phase space distributed events. This approach gives also a glimpse into the future of high energy physics, where experiments publish new type of measurements in a radically multidimensional way.


\end{abstract}

\pacs{Valid PACS appear here}
\maketitle


High energy physics data needs to be corrected for experimental effects induced by kinematically non-uniform response of the detector and reconstruction algorithms. These corrections are usually dubbed under efficiency times acceptance corrections, implemented often by dividing the measured number of background corrected events in a histogram bin with the efficiency value obtained by simulations. In addition, unfolding of histograms of observables is often done to account for distortions of resolution (variance) and absolute value (bias) induced by the detector. Unfolding is usually treated with a stochastic smearing matrix, obtained via simulations. It is the art of unfolding and optimization algorithms, with suitable regularization and possible additional physical boundary conditions, which then ``inverts'' this response matrix in an algebraic or probabilistic way. An important part is the uncertainty estimation, both in purely statistical terms but also in systematic ways. Well known is that many of the naive bin-by-bin methods are severaly biased towards ``Monte Carlo truth'' \cite{cowan1998statistical}.

A crucial step of the measurement is the proper definition of the fiducial phase space. Ideally, this should be defined in terms of proper final state observables, such as transverse momentum and pseudorapidity of charged particles, within geometrically visible and electrically active detector volume. A fiducial measurement, by its very definition, minimizes extrapolations of data and thus also minimizes the model dependence. In this work, we go further and propose \textit{maximally} Monte Carlo event generator independent way to implement fiducial efficiency corrections. For the best of our knowledge, this philosophy and approach what we call \textit{DeepEfficiency}, based on Deep Neural Networks (DNN), is a new one. However, in other context the networks are well known in high energy physics, especially in signal-background separation. For a recent review see \cite{guest2018deep}, or for the pioneering studies \cite{denby1988neural}.

For concreteness, let us define our observable degrees of freedom at the level of individual final state particles. Our final state consist of $N$ charged particles (tracks) with 3-momentum $\vec{p}$ of each being measured. This final state spans a real valued vector space $\mathbb{R}^{3N}$, from which one can construct a set of Lorentz scalars and other non-scalar observable distributions, such as transverse momenta or angular ones. The detector has a finite probability $\mathbb{P}$ of seeing the $N$-track event, given the trigger and tracking efficiencies. This expected probability is represented with an efficiency mapping $\mathcal{E} : \mathbb{R}^{3N} \rightarrow \mathbb{P}$. Clearly, by using space-time symmetries of interest, one can often do dimensional reduction. However, treating maximally the abstract detector space, forbids us from doing so. That is, we want to take into account all the correlations that the detector may induce in terms of efficiency losses.

A simplification, dimensional reduction but also a loss of information would occur, if one would for example factorize all tracks and get the full event detection efficiency as $\mathcal{E}_{tot} = \mathcal{E}_1 \mathcal{E}_2 \cdots \mathcal{E}_N$. However, for very high dimensional problems, one may do this. We point out here that the efficiency function $\mathcal{E}$ can be used also within the context of Matrix Element Method (MEM) type likelihood analyses, in addition to the fiducial measurements which are of our interest here. For typical approximations within MEM applications, see \cite{fiedler2010matrix}.

What we simply now do is learn this high dimensional detector efficiency function $\mathcal{E}$ using a fully connected, multilayer DNN in a regression mode. That is, we simply use the neural network as a high dimensional function approximator and interpolator - as a mathematical hammer. The exact architecture, the cost function and its regularization and the gradient descent methods are art forms each of their own, we return to these in extended descriptions. Because we use the fully differential observable kinematic information of the final states, we can even use a pure phase space Monte Carlo event generator as an input to the detector simulation. Perhaps the most intuitive proof of the vanishing dependence on the event generator is based on a multidimensional histogram division picture with hyperbin volumes approaching zero.

A practical requirement for the generator is that a large enough event statistics is produced continuously for all corners of the phase space of interest, naturally. For final states containing jets, a pure phase space Monte Carlo is not feasible. Thus in practise one proceeds as usual with a realistic MC generator of interest. One may do this also in other cases, naturally. Here we do not touch the highly interesting but complex topic of matching optimally parton level, hadron level and detector level objects and sub-structures, which is basically always non-trivial, sometimes ill-posed, with hard or soft QCD processes.

In optimizing (training) the network parameter set $\Theta$, we minimize the so-called cross entropy cost function suitable for our probabilistic inference problem
\begin{equation}
L(\Theta, \mathcal{S}) = -\frac{1}{|\mathcal{S}|}\sum_{i \in \mathcal{S}} \left[ R_i \ln \mathcal{E}_i + (1 - R_i) \ln (1 - \mathcal{E}_i) \right],
\end{equation}
where the sum is over all events in the simulation sample $\mathcal{S}$. The known response $R$ is 0 for efficiency lost events, and 1 for selected events, both within the fiducial phase space. That is, one must not include events outside the fiducial phase space in the sample, by construction. A cutoff regularization is used for the rare $\mathcal{E} \rightarrow 0$ singularity. The word entropy comes from the functional form of the cost, related to the Kullback-Leibler \cite{kullback1951information} divergence which is a difference between cross entropy and entropy, and the KL-divergence itself is related to the Maximum Likelihood principle. In addition, one can add typical regularization schemes such as $\ell_1$- (sparsity, Laplace) or $\ell_2$-norm (smoothness, Gaussian) based, in order to avoid overfitting. This is avoided also effectively by choosing the minimal network architecture which results in efficiency corrections with minimal bias and variance on simulated samples. Also the training sample size needs to be high enough, given that the deep network architectures can contain $\mathcal{O}(10^4-10^6)$ free parameters.

After training, the efficiency inversion of data and arbitrary observables of interest are obtained with
\begin{equation}
\frac{d N}{d \mathcal{O}} = h_\mathcal{O}( \{ \vec{p} \}) \odot [\mathcal{E}( \{ \vec{p} \})]^{-1},
\end{equation}
where $h_\mathcal{O}$ is a probability distribution estimator operator, typically a bin width $\Delta \mathcal{O}$ normalized histogram. That is, the discretization of observables only enters at this point. The point-wise operator $\odot$ is defined as an integral (sum) over the event sample and the weight $\mathcal{E}( \{ \vec{p} \} )$ is obtained from the neural network. Simply put, event-by-event, one constructs the observable of interest and calls a weighted histogram fill with a weight $\mathcal{E}^{-1}$. Thus, in an extraordinary smooth way, one can efficiency correct \textit{simultaneously} arbitrary number of single or multidimensional histograms of observables using the same weight. For a related mathematical discussion, see the Horvitz-Thompson unbiased estimator \cite{horvitz1952generalization}. Differential cross sections are obtained in a standard way normalizing by integrated luminosity. A straightforward way to obtain statistical uncertainties is via Efron's bootstrap re-sampling. The possible algorithmic (network) bias should be empirically tested using simulations, observable by observable, which is easily automated.

The performance was numerically evaluated using the full ALICE detector simulation with input being low-mass, low-$p_T$ central exclusive QCD diffraction, which is a prominent ``glueball production process'' at the LHC, decaying in this case to charged meson final states $(\pi^+\pi^-, K^+K^-)$. We obtained solid results with a 5-layer network with $\sim100$ neurons per layer with a 6-dimensional input, using hyperbolic tangent activation functions and a final layer with one sigmoidal output function. The inversion performance in terms of $\chi^2 / bin$ was close to unity for a variate of one and two dimensional differential distributions, also in the steep tails. TensorFlow 1.9 \cite{abadi2016tensorflow} was used as the computational framework. The event generator independence lemma was observed to hold with Monte Carlo samples driven by different scattering amplitudes, such as photoproduction of $\rho^0 \rightarrow \pi^+\pi^-$ versus $\pi^+\pi^-$ production via double Pomeron exchange. A typical efficiency correction bias coming from unknown angular dependence or spin polarization densities, due to uncertain non-perturbative scattering amplitudes for both continuum and resonance production - is a solved problem now. Also the efficiency correction bias, propagating from a priori unknown but parametrized non-perturbative proton elastic form factors and inelastic structure functions driving the system soft transverse momentum distributions, vanishes.

Summarizing, we introduced a formally optimal and generic efficiency inversion algorithm for fiducial measurements. As a reminder, fiducial measurements are measurements minimizing the kinematic-geometrical acceptance extrapolations. However, in certain cases non-fiducial extrapolations are needed, for example the case of a fully angular flat phase space definition requirement for spin polarization studies. Another interesting topic is the high dimensional, continuous unfolding of distributions, which we did not discuss here. However, one can fuse single dimensional matrix unfolding algorithms with the efficiency inversion of DeepEfficiency.

A proof-of-principle Python/C++ code including the full algorithmic chain, is available under the MIT license at \href{https://github.com/mieskolainen/deepefficiency}{github.com/mieskolainen/deepefficiency}. Also, \href{https://opendata.cern.ch}{opendata.cern.ch} provides hundreds of simulated and real datasets from ATLAS and CMS for possible test input. Acknowledgements: Risto Orava is thanked for discussions and Evgeny Kryshen for ALICE simulation resources.

\bibliography{apssamp}

\end{document}